\documentclass[aps,prb,twocolumn,groupedaddress,noshowpacs,checklength,amsmath]{revtex4-1}

\usepackage{graphicx}
\usepackage{amssymb}
\usepackage{amsmath}
\usepackage{cancel}

\newcommand{\grad}{\mbox{\boldmath $\nabla$}}

\def\ihat{{\bf \hat x}}
\def\jhat{{\bf \hat y}}
\def\khat{{\bf \hat z}}

\begin{document}

\title{The torque on a dipole in uniform motion}

\author{David J.~Griffiths}\email[Electronic address: ]{griffith@reed.edu}
\affiliation{Department of Physics, Reed College, Portland, Oregon  97202}
\author{V.~Hnizdo}
\affiliation{National Institute for Occupational Safety and Health, Morgantown, West Virginia 26505}

\begin{abstract}
We calculate the torque on an ideal (point) dipole moving with constant velocity through uniform electric and magnetic fields.
\end{abstract}

\maketitle

\section{  Introduction}

The torque on an electric dipole {\bf p}, at rest in a uniform electric field {\bf E}, is
\begin{equation}
{\bf N}= {\bf p}\times{\bf E}.
\end{equation}
The torque on a magnetic dipole {\bf m}, at rest in a uniform magnetic field {\bf B}, is
\begin{equation}
{\bf N}= {\bf m}\times{\bf B}.
\end{equation}
But what if the dipole is moving, at a constant velocity {\bf v}?  It is well known that a moving electric dipole acquires a {\it magnetic} dipole moment \cite{VH}
\begin{equation}
{\bf m}_v =-({\bf v}\times {\bf p}),
\end{equation}
so one might guess that the torque on a moving electric dipole in uniform electric and magnetic fields would be \cite{Ref}
\begin{equation}
{\bf N} = ({\bf p}\times{\bf E})+({\bf m}_v\times{\bf B}).
\end{equation}
Similarly, a moving magnetic dipole acquires an {\it electric} dipole moment \cite{acqp}
\begin{equation}
{\bf p}_v = \frac{1}{c^2}({\bf v}\times {\bf m}),
\end{equation}
and one might guess that the torque on a moving magnetic dipole would be
\begin{equation}
{\bf N} = ({\bf m}\times{\bf B})+({\bf p}_v\times{\bf E}).
\end{equation}
But these formulas are incorrect; in each case there is a third term:\cite{MP}
\begin{eqnarray}
{\bf N}_p &=& ({\bf p}\times{\bf E})+({\bf m}_v\times{\bf B})+{\bf v} \times ({\bf p}\times {\bf B}),\\
{\bf N}_m &=& ({\bf m}\times{\bf B})+({\bf p}_v\times{\bf E}) + {\bf v}\times({\bf p}_v\times {\bf B}).
\end{eqnarray}
In general, then,
\begin{equation}
{\bf N} = ({\bf p}\times{\bf E})+({\bf m}\times{\bf B})+{\bf v} \times ({\bf p}\times {\bf B}).
\end{equation}
Our purpose in this note is to derive that result. \cite{NU}
\bigskip

\vskip0in
\begin{figure}[t]
\hskip-.1in\scalebox{.7}[.7]{\includegraphics{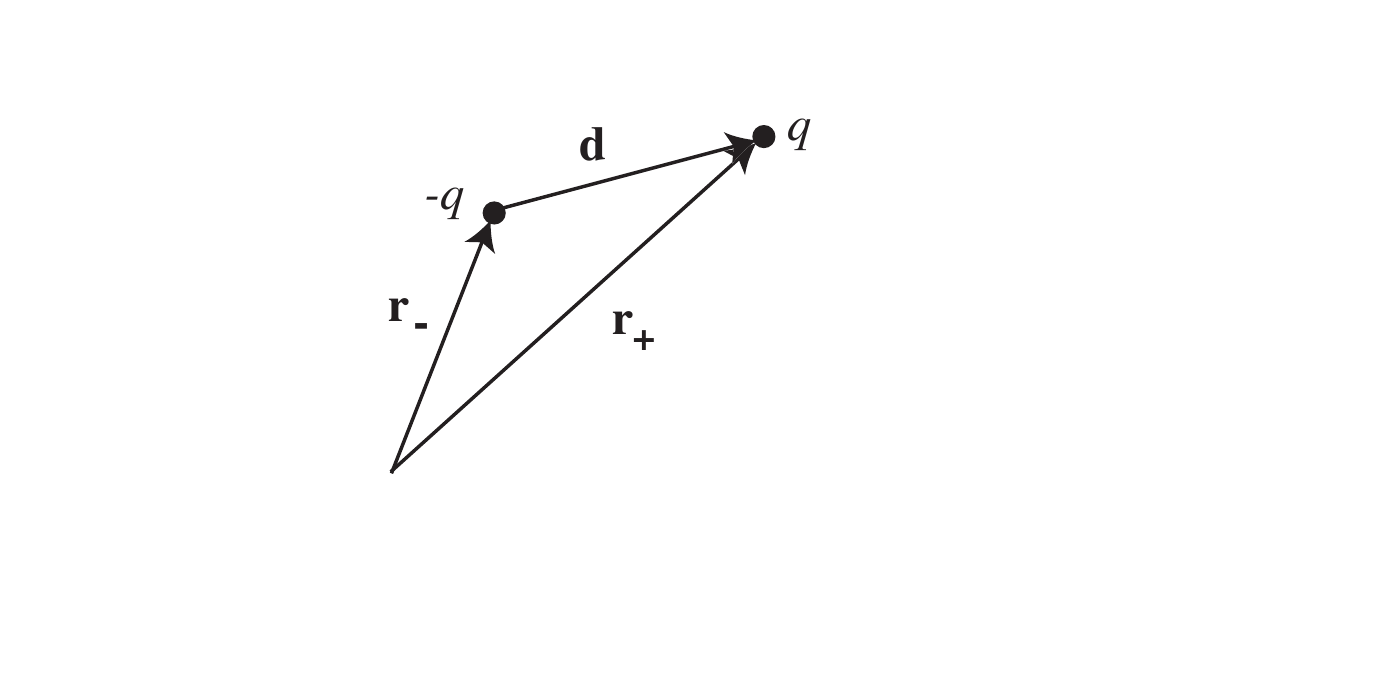}}
\vskip-.5in
\caption{Electric dipole.}
\end{figure}

\section{ Electric Dipole in Motion}

An electric dipole consists of positive and negative charges $\pm q$ separated by a displacement ${\bf d}$ (Figure 1).  The force on a point charge is
\begin{equation}
{\bf F} = q\left({\bf E} + {\bf v}\times {\bf B}\right),
\end{equation}
so the torque \cite{origin} on a dipole of moment ${\bf p}=q{\bf d}$ is
\begin{eqnarray}
{\bf N} &=& ({\bf r}_+\times {\bf F}_+)+({\bf r}_-\times {\bf F}_-)\nonumber\\
&=&({\bf r}_+-{\bf r}_-)\times \left[q\left({\bf E}+{\bf v}\times {\bf B}\right)\right]\nonumber\\
&=&q{\bf d}\times\left({\bf E}+{\bf v}\times {\bf B}\right)\nonumber\\
&=& ({\bf p}\times {\bf E})+{\bf p}\times({\bf v}\times {\bf B}).
\end{eqnarray}
Using the vector identity ${\bf p}\times({\bf v}\times {\bf B})= {\bf v}\times({\bf p}\times {\bf B})+({\bf p}\times{\bf v})\times {\bf B}$, the previous result becomes
\begin{equation}
{\bf N} = ({\bf p}\times {\bf E})+({\bf m}_v\times{\bf B})+{\bf v}\times({\bf p}\times{\bf B}).
\end{equation}

That is surely the {\it easiest} way to obtain Eq.~(7) (it borrows from an argument by V.~Namias\cite{Namias}).  But the magnetic analog is not so straightforward,\cite{mag} so let's repeat the derivation, this time treating the electric dipole as the point limit of a uniformly polarized object.\cite{Man}   Polarization ({\bf P}) and magnetization ({\bf M}) together constitute an antisymmetric second-rank tensor:
\begin{equation}
P^{\mu\nu} = \begin{pmatrix}0&cP_x&cP_y&cP_z\\-cP_x&0&-M_z&M_y\\-cP_y&M_z&0&-M_x\\-cP_z&-M_y&M_x&0
\end{pmatrix},
\end{equation}
and the transformation rule gives\cite{DJG1}
\begin{eqnarray}
&&P_z{=}P_z',\ P_x{=}\gamma(P_x'{-}\frac{v}{c^2}M_y'),\ P_y{=}\gamma(P_y'{+}\frac{v}{c^2}M_x'),\\
&&M_z{=}M_z',\ M_x{=}\gamma(M_x'{+}vP_y'),\ M_y{=}\gamma(M_y'{-}vP_x').
\end{eqnarray}
We use primes for the ``proper" (rest) system of the dipole (${\cal S'}$); no prime means the ``lab" frame (${\cal S}$), through which the dipole is moving with velocity $v\,\khat$.  Notice that if ${\bf M}'=0$, then ${\bf M}=-({\bf v} \times {\bf P})$, confirming Eq.~(3), and if ${\bf P}'=0$, then ${\bf P}=(1/c^2)({\bf v}\times {\bf M})$, confirming Eq.~(5).

Now, suppose we have an electric dipole ${\bf p}_0$ at rest at the origin in ${\cal S}'$:
\begin{equation}
{\bf P}'(x',y',z')= {\bf p}_0\delta(x')\delta(y')\delta(z').
\end{equation}
Assume first that ${\bf p}_0$ is perpendicular to {\bf v}, say
\begin{equation}
{\bf p}_0=p_0\,\ihat.
\end{equation}
Then
\begin{eqnarray}
P_x&=&\gamma p_0\delta(x')\delta(y')\delta(z')= \gamma p_0\delta(x)\delta(y)\delta(\gamma(z-vt))\nonumber\\
&=& p_0\delta(x)\delta(y)\delta(z-vt),\\
M_y&=&-v\gamma p_0\delta(x')\delta(y')\delta(z')\nonumber\\
&=&-v p_0\delta(x)\delta(y)\delta(z-vt),
\end{eqnarray}
and all other components are zero.  According to the Lorentz law, the force density is
\begin{equation}
{\bf f} = \rho{\bf E} + {\bf J}\times {\bf B},
\end{equation}
where \cite{DJG2}
\begin{equation}
\rho = -\grad \cdot {\bf P},\quad {\bf J}=\grad\times {\bf M} + \frac{\partial {\bf P}}{\partial t}.
\end{equation}
In our case,\cite{diff}
\begin{equation}
\rho = -p_0\delta'(x)\delta(y)\delta(z-vt),
\end{equation}
\begin{equation}
\grad{\times} {\bf M} {=} vp_0\left[\delta(x)\delta(y)\delta'(z{-}vt)\,\ihat {-}\delta'(x)\delta(y)\delta(z{-}vt)\,\khat\right],
\end{equation}
\begin{equation}
\frac{\partial \bf P}{\partial t}= -vp_0\delta(x)\delta(y)\delta'(z-vt)\,\ihat,
\end{equation}
so
\begin{equation}
{\bf J} = -vp_0 \delta'(x)\delta(y)\delta(z-vt)\,\khat=-p_0 \delta'(x)\delta(y)\delta(z-vt)\,{\bf v}
\end{equation}
and
\begin{equation}
{\bf f} = -p_0\delta'(x)\delta(y)\delta(z-vt)[{\bf E} + ({\bf v}\times {\bf B})].
\end{equation}
The torque is
\begin{eqnarray}
{\bf N} &=&\int({\bf r}\times {\bf f})\,d^3{\bf r}\nonumber\\
&=&-\,p_0\left\{\int \delta'(x)\delta(y)\delta(z-vt){\bf r}\,dx\,dy\,dz\right\}\nonumber\\
&&\times [{\bf E} + ({\bf v}\times {\bf B})],
\end{eqnarray}
where ${\bf r} = (x,y,z)$.  The $y$ and $z$ components of the integral are zero; the $x$ component is \cite{dbp}
\begin{equation}
\int x \delta'(x)\delta(y)\delta(z-vt)\,dx\,dy\,dz= \int x\delta'(x)\,dx = -1,
\end{equation}
so
\begin{equation}
{\bf N} = p_0\,\ihat \times [{\bf E} + ({\bf v}\times {\bf B})]= ({\bf p}_0\times {\bf E}) + {\bf p}_0 \times ({\bf v} \times {\bf B}),
\end{equation}
which agrees with Eq.~(11), and hence confirms Eq.~(7) for the case ${\bf p}_0$ perpendicular to {\bf v}.  (For this orientation the electric dipole moments in ${\cal S}$ and ${\cal S}'$ are the same---there is no Lorentz contraction.  This can be confirmed by integrating Eq.~(18): ${\bf p} \equiv \int {\bf P}\, d^3{\bf r}={\bf p}_0$.)

If ${\bf p}_0$ is parallel to {\bf v},
\begin{equation}
{\bf p}_0=p_0\,\khat,
\end{equation}
giving
\begin{equation}
P_z=p_0\delta(x')\delta(y')\delta(z')= \frac{p_0}{\gamma}\delta(x)\delta(y)\delta(z-vt),
\end{equation}
with all other components of {\bf P} and {\bf M} being zero.
In this case,
\begin{equation}
\rho = -\frac{p_0}{\gamma}\delta(x)\delta(y)\delta'(z-vt)
\end{equation}
and
\begin{equation}
{\bf J} = -\frac{vp_0}{\gamma}\delta(x)\delta(y)\delta'(z-vt)\,\khat ,
\end{equation}
so that
\begin{equation}
{\bf f} = -\frac{p_0}{\gamma}\delta(x)\delta(y)\delta'(z-vt)[{\bf E} + ({\bf v}\times {\bf B})].
\end{equation}
The torque is
\begin{equation}
{\bf N} =\frac{{\bf p}_0}{\gamma}\times [{\bf E} + ({\bf v}\times {\bf B})].
\end{equation}
Because of Lorentz contraction, ${\bf p} = {\bf p}_0/\gamma$---as is confirmed by integrating Eq.~(31)---and again we recover Eq.~(7).

\section{ Magnetic Dipole in Motion}

Now, suppose we have a magnetic dipole ${\bf m}_0$, at rest at the origin, in ${\cal S}'$:
\begin{equation}
{\bf M}'(x',y',z')= {\bf m}_0\delta(x')\delta(y')\delta(z').
\end{equation}
Assume first that ${\bf m}_0$ is perpendicular to {\bf v}, say
\begin{equation}
{\bf m}_0=m_0\,\ihat.
\end{equation}
From the transformation rule,
\begin{eqnarray}
&&P_y{=}\gamma \frac {vm_0}{c^2}\delta(x')\delta(y')\delta(z'){= }\frac{vm_0}{c^2}\delta(x)\delta(y)\delta(z{-}vt),\\
&&M_x{=}\gamma m_0\delta(x')\delta(y')\delta(z'){=}m_0 \delta(x)\delta(y)\delta({z-}vt).
\end{eqnarray}
This time
\begin{equation}
\rho =-\grad\cdot {\bf P}= -\frac {vm_0}{c^2}\delta(x)\delta'(y)\delta(z-vt),
\end{equation}
\begin{equation}
\grad{\times} {\bf M}{=} m_0\left[\delta(x)\delta(y)\delta'(z{-}vt)\,\jhat {-}\delta(x)\delta'(y)\delta(z{-}vt)\,\khat\right],
\end{equation}
\begin{equation}
\frac{\partial {\bf P}}{\partial t}=-\frac{v^2}{c^2}m_0\delta(x)\delta(y)\delta'(z-vt)\,\jhat,
\end{equation}
so
\begin{equation}
{\bf J} = m_0\left[\frac{1}{\gamma^2}\delta(x)\delta(y)\delta'(z-vt)\jhat - \delta(x)\delta'(y)\delta(z-vt)\khat\right],
\end{equation}
and
\begin{eqnarray}
{\bf f} &=& -\frac{vm_0}{c^2}\delta(x)\delta'(y)\delta(z-vt){\bf E}\nonumber\\
&&+\, m_0\bigl[\frac{1}{\gamma^2}\delta(x)\delta(y)\delta'(z-vt)\jhat
 -\ \delta(x)\delta'(y)\delta(z-vt)\khat\bigr]\nonumber\\
 &&\quad\quad\quad\times {\bf B}.
\end{eqnarray}
The torque is
\begin{eqnarray}
{\bf N} &=& \frac{vm_0}{c^2}(\jhat\times {\bf E})-\frac{m_0}{\gamma^2}\,\khat\times(\jhat \times {\bf B}) + m_0\,\jhat\times (\khat\times {\bf B})\nonumber\\
&=&({\bf p}_v\times {\bf E}) +({\bf m}_0\times {\bf B})+{\bf v}\times ({\bf p}_v\times {\bf B}),
\end{eqnarray}
in agreement with Eq.~(8), for ${\bf m}_0$ perpendicular to {\bf v}.  (In this orientation, the magnetic dipole moments in ${\cal S}$ and ${\cal S}'$ are the same, as we confirm by integrating Eq.~(39): ${\bf m} \equiv \int {\bf M}\, d^3{\bf r}={\bf m}_0$.)

If ${\bf m}_0$ is parallel to {\bf v},
\begin{equation}
{\bf m}_0=m_0\,\khat,
\end{equation}
giving
\begin{equation}
M_z=m_0\delta(x')\delta(y')\delta(z')= \frac{m_0}{\gamma}\delta(x)\delta(y)\delta(z-vt),
\end{equation}
with all other components of {\bf P} and {\bf M} being zero.
In this case $\rho$ and $\partial {\bf P}/\partial t$ vanish, leaving
\begin{equation}
{\bf J}= \frac{m_0}{\gamma}\left[\delta(x)\delta'(y)\delta(z-vt)\,\ihat-\delta'(x)\delta(y)\delta(z-vt)\,\jhat\right],
\end{equation}
so
\begin{equation}
{\bf f} = \frac{m_0}{\gamma}\left[\delta(x)\delta'(y)\delta(z-vt)\,\ihat-\delta'(x)\delta(y)\delta(z-vt)\,\jhat\right]\times{\bf B},
\end{equation}
and the torque is
\begin{equation}
{\bf N} =\frac{1}{\gamma}({\bf m}_0\times {\bf B}),
\end{equation}
which is again consistent with Eq.~(8) because in this orientation ${\bf p}_v=0$ and, integrating Eq.~(47), ${\bf m} = {\bf m}_0/\gamma$.

\section{Hidden Momentum}

In the presence of an electric field, a magnetic dipole (a loop of electric current) carries ``hidden momentum"\cite{HM} ${\bf p}_h=(1/c^2)({\bf m}\times{\bf E})$ and hence also hidden momentum
\begin{equation}
{\bf L}_h=\frac{1}{c^2}{\bf r}\times ({\bf m}\times {\bf E}).
\end{equation}
It is ``hidden" in the sense that it is not associated with overt motion of the object as a whole; it occurs in systems with internally moving parts, such as current loops.\cite{HMM} The {\it total} angular momentum is the sum of ``overt" and hidden components:
\begin{equation}
{\bf L} = {\bf L}_o+{\bf L}_h,
\end{equation}
and the torque is its rate of change:\cite{mdot}
\begin{eqnarray}
{\bf N} &=& \frac{d{\bf L}_o}{dt} + \frac{d{\bf L}_h}{dt} \nonumber \\
&=& {\bf N}_o + \frac{1}{c^2}{\bf v}\times({\bf m}\times {\bf E}).
\end{eqnarray}
Thus, combining Eqs.\ (9) and (53), the ``overt" torque on a moving dipole is
\begin{equation}
{\bf N}_o=({\bf p} \times {\bf E}) +({\bf m}\times{\bf B})+{\bf v}\times({\bf p}\times {\bf B})-\frac{1}{c^2}{\bf v}\times({\bf m}\times {\bf E}).
\end{equation}

By contrast, a (stationary) {\it electric} dipole has no internally moving parts, and it harbors no hidden momentum.\cite{d.rest} However, when it is in {\it motion} it picks up a magnetic dipole moment ${\bf m}_v=-({\bf v})\times{\bf p}$ [Eq.\ (3)] that {\it does} carry hidden momentum,\cite{circ.current} so Eq.\ (54) (with ${\bf m}={\bf m}_v$) applies to this case as well.\cite{KH}

\section*{Acknowledgment}
V.H. co-authored this note in his private capacity; no official support or endorsement by the Centers for Disease Control and Prevention is intended or should be inferred.


\begin{thebibliography}{99}
\bibitem{Ref} In this paper, except for Eqs.\ (1) and (2), {\bf p} and {\bf m} are the dipole moments of the {\it moving} dipoles---{\it not} the ``proper" moments in the dipoles' rest frame.
\bibitem{origin} We shall calculate all torques and angular momenta with respect to the origin.
\bibitem{diff} When attached to a delta function, the prime denotes differentiation with respect to its argument.
\bibitem{HMM} Hidden momentum is mechanical in nature, even though it arises (typically) in electromagnetic contexts. For details see Ref.\ 4, or Ref.\ 10 (Example 12.13).
\bibitem{circ.current} Note that a moving electric dipole carries a component of circulating current; see Hnizdo, Ref.\ 1.
\end{thebibliography}
\end{document}